\documentclass[onecolumn,showpacs,amsmath,amssymb,prd,nofootinbib]{revtex4-2}
\usepackage{graphicx}
\usepackage{epsfig}
\usepackage{enumerate}
\usepackage{bm}
\usepackage{amsbsy}
\def\sss{\scriptscriptstyle}
\def\^#1{^{\sss #1}}
\def\_#1{_{\sss #1}}
\def\av#1{\langle #1\rangle}
\def\avt#1{\langle #1\rangle_t}
\def\avm#1{\langle #1\rangle_m}
\def\beq{\begin{equation}}
\def\eeqno#1{\label{#1}\end{equation}}

\def\az{a_{\scriptscriptstyle 0}}
\def\A{\textbf{\textsc{A}}}

\def\U{\mathcal{U}}
\def\I{\mathcal{I}}
\def\J{\mathcal{J}}

\def\Q{\mathcal{Q}}
\def\L{\mathcal{L}}

\def\N{\mathcal{N}}
\def\F{\mathcal{F}}

\def\azg{\mathcal{A}_0}

\def\rar{\rightarrow}
\def\s{\sigma}
\def\r{\rho}
\def\l{\lambda}

\def\drt{d^3r}
\def\f{\phi}
\def\grad{\bm{\nabla}}
\def\div{\bm{\nabla}\cdot}
\def\gf{\grad\phi}
\def\gfN{\grad \f_N}

\def\vv{\textbf{v}}
\def\va{\textbf{a}}
\def\vg{\textbf{g}}

\def\fN{\f\_N}

\def\m{\mu}
\def\a{\alpha}
\def\b{\beta}

\def\d{\delta}
\def\z{\zeta}

\def\n{\nu}

\def\gN{g_{\scriptscriptstyle N}}
\def\vgN{\vg_{\scriptscriptstyle N}}

\def\eff{\bm{\Im}}

\newcommand{\vvr}{ \textbf{r}}

\def\rM{r\_M}

\def\IK{\I\_K}
\def\Lf{L_f}

\begin{document}
\title{Mixed modified-gravity--modified-inertia MOND theories}

\author{Mordehai Milgrom}
\affiliation{Department of Particle Physics and Astrophysics, Weizmann Institute}

\begin{abstract}
It is plausible that in a more fundamental theory of MOND both gravity and inertia will differ from their Newtonian versions. Such theories, if they embody the basic tenets of MOND, will share its primary predictions with existing theories, which modify either gravity or inertia. But they are expected to differ in their structure and in their secondary predictions.
Here, I discuss families of nonrelativistic MOND theories that modify both the kinetic and the gravitational actions of Newtonian dynamics. Each family is spanned by one parameter $0\le\eta\le 1$, which is the scaling dimension of the deep-MOND limit of all terms in the action.
Pure modified gravity ($\eta=0$) and pure modified inertia ($\eta=1$) are the end cases in each family. I speak of families, since the modified action terms themselves are not strictly specified, but are treated in rather general terms. For example, for $\eta=0$ we do not get a specific modified-gravity MOND theory, but a large class of such theories.
These theories are artificial in that they are not based on a single unifying idea that dictates the modification of the two components. But they are still useful in pinpointing some idiosyncracies of such theories. For example, neither of the two modified terms, by itself, leads to MOND phenomenology; only their combined effect does. I show how the two combine to make some of the primary MOND predictions, such as asymptotic flatness of rotation curves, the mass-asymptotic-speed relation, and the deep-MOND, virial, mass-velocity-dispersion ($M-\sigma$) relation. Another idiosyncracy of such theories is the emergence of different interpolating functions in the kinetic equations of motion and in the gravitational field equations, each relevant under different circumstances, and neither of which have the standard deep-MOND limit.
Such examples alert us to the wide range of theories -- nonrelativistic and relativistic -- that can lead to MOND phenomenology.

\end{abstract}
\maketitle

\section{introduction}
MOND theories aim primarily to account for the dynamics of systems governed by gravity. In particular, systems -- such as galactic systems and the Universe at large -- that evince mass discrepancies. Accounting for these dynamics with Newtonian dynamics (ND) and general relativity requires the invocation of dark matter, while a MOND theory strives to do it without dark matter.
The basic tenets of MOND are listed, e.g., in the recent Ref. \cite{milgrom25}.
Reviews of MOND that describe its basics, its predictions and their confrontation with observations, and the various underlying theories that have been propounded to date can be found in Refs. \cite{fm12,milgrom14,milgrom20a,mcgaugh20,merritt20,bz22,fd25}; in particular, \cite{lelli26} is an extensive ``white paper'' on the MOND research program.
\par
A basic starting point of nonrelativistic (NR) MOND theories is the NR Newtonian action that describes gravitating systems of pointlike masses, $m_i$, whose trajectories are $\vvr_i(t)$
 \beq  I=-\frac{1}{8\pi G} \int dt~\drt ~(\gfN)^2 -\sum_i m_i\int dt~\fN(\vvr_i(t))+\frac{1}{2}\sum_i m_i\int dt~\dot\vvr_i^2,     \eeqno{newtaction}
 where $\fN$ is the Newtonian gravitational potential.
In the continuum-limit version of this action, which I shall use occasionally,
$\sum_i m_i$ is replaced by  $\int \drt \r$, where $\r(\vvr,t)=\sum_i m_i\d^3[\vvr-\vvr_i(t)]$.
\par
It is possible that a fundamental theory underlying MOND, a FUNDAMOND, will not be based on an action principle.
But, action formulations have their advantages; so, at the most basic NR level, we want to generalize this Newtonian action to one that satisfies the basic tenets of MOND.
\par
The workhorses NR MOND theories, AQUAL \cite{bm84} (standing for aquadratic Lagrangian) and QUMOND \cite{milgrom10} (standing for quasilinear MOND), as well as TRIMOND \cite{milgrom23b} (standing for tripotential MOND), and the class of more general MOND theories described in Ref. \cite{milgrom14b}, modify only the first term in the above Newtonian action (\ref{newtaction}) -- i.e., the gravitational, Poisson term; they are pure modified-gravity (MG) theories. The MOND theories discussed in \cite{milgrom94}, on the other hand, modify only the last, inertial term in the Newtonian action; they are pure modified-inertia (MI) theories. There have also been described MI theories that modify directly the kinetic particle equations of motion (EoMs), without an action basis \cite{milgrom22a,milgrom23}. The same distinction characterizes existing relativistic MOND theories (see Ref. \cite{lelli26} and references therein), which can all be characterized as MG theories: Their matter actions still have the standard form with matter degrees of freedom (DoFs) coupling minimally to some metric, but this metric is not determined, as it does in general relativity, from the Einstein field equations, which, in turn, derive from the Einstein-Hilbert action.
\par
While on consistency grounds alone, pure MG is acceptable on its own, pure MI is less so. It stands to reason that if we are to modify inertia, as encapsulated in the matter actions, then -- at least at the relativistic level -- gravity should also be modified. This is because like ``matter'', relativistic gravity too has inertia, which is encapsulated in the gravitational action. Thus, it is conceivable that a more fundamental MOND theory, or even an effective MOND theory that emerges from more fundamental considerations, will involve modifications of both gravity and inertia.
\par
Harking back to important precedents, special relativity (SR) brought about modification of Newtonian inertia, without modifying Newtonian gravity. It had, however, been clear from the outset that Newtonian gravity is at odds with the principles of SR, and that it must also be modified to be compatible with SR, as indeed has been realized in general relativity.
SR replaced the Galilean boost invariance underlying Newtonian inertia, with the Lorentz boost invariance. It may well be that a new symmetry will also turn out to underlie MOND dynamics, either only in the deep-MOND regime, or for the full acceleration range, with $\az$ entering, as $c$ enters Lorentz transformations \cite{milgrom06a}.
\par
Still within standard, pre-MOND dynamics, today we are in a situation where classical particle and field inertia (as encapsulated in the matter action) has been modified to ``quantum inertia'' -- as per Schr\"{o}dinger's and Dirac's equations and quantum field theory, but gravity is still waiting to be convincingly incorporated in the quantum framework.
\par
The anomalies that motivated these modifications concerned field and particle dynamics, without, at first, concerning or involving gravity.\footnote{If we discount the residual perihelion shift of Mercury, which was known, but is not thought to have acted as a catalyst in this regard.} So, it was natural and necessary that they were addressed first by modifying the dynamics of particle and the electromagnetic field.
\par
In contradistinction, the anomalies that have led to MOND involve gravity in a major way. So much so, that it has turned out to be enough to modify gravity alone to reproduce the most salient aspects of MOND phenomenology.
Since this approach seems more straightforward and amenable to progress we are in a situation where practically all full-fledged MOND theories, relativistic and NR, are of the MG type.
However, modifying inertia to account for MOND phenomenology has its matter-of-principle attractions \cite{milgrom22a,milgrom23}, and furthermore, MI may make rather different predictions from MG concerning secondary phenomena \cite{milgrom22a,milgrom23}, and so it may be forced on us by observations and experiments.
\par
In the MOND context, in either MG or MI theories, it is a single term in the action that is responsible for the deep-MOND behavior, and for the transition from the Newtonian to the deep-MOND limit (DML). In the mixed theories these aspects rest on the two terms, and we need to demonstrate the scheme that is responsible for this MOND behaviors. It is also interesting to gain some notion of how such theories may differ in their predictions from existing MOND theories.
\par
In Ref. \cite{milgrom25a}, I presented two specific families of DML, toy models that combine MG and MI. These are however rather limited in scope:
They are both strict DML modeled with no visible possibility to connect them with the Newtonian regime. They also both involve only a single potential as the gravitational DoF, $\psi$.
Furthermore, in one family gravity is forced to be strictly linear, as in Newtonian gravity. While these are useful toy models, they are too restricted.
\par
Here I consider mixed MG-MI theories in rather more general terms.
It is expected that in a MOND theory -- even if only an effective approximation of some more fundamental theory -- the departure from standard dynamics of both gravity and inertia would result from the same mechanism, for example, from the interaction of matter and gravity with some omnipresent agent whose dynamics remains behind the scene.
\par
I cannot offer such a desirable unifying origin for the theories I discuss here.
Rather, their modifications of the gravitational and kinetic actions are treated as independent, with the proviso that their combined effects give equations of motion that obey the basic tenets of MOND.
Yet, such model theories can be useful as perhaps pointing the way to better-motivated theories. They are also useful heuristic tools that can give some grasp of the possible range of MOND theories and highlight the possible diversity of secondary predictions regarding certain phenomena.
\par
Indeed, part of my motivation for propounding these theories  -- along with other model theories discussed in Refs.  \cite{milgrom22a,milgrom23,milgrom23b,milgrom25a} -- is to bring home the point that one should not identify MOND with AQUAL and QUMOND. While these had been the only full-fledged, NR MOND theories around for many years, and have serves us very well, one should not take all their predictions to be absolute predictions of MOND.
\par
Section \ref{generalities} discusses some preliminary generalities regarding MG, MI, and the possible interplay between them. In Sec. \ref{theories}, I present the general modified actions. Section \ref{dmlsec} discusses their behavior in the DML. Section \ref{rotasym} discusses additional aspects of the theories, such as the asymptotic potential field, and the resulting asymptotic rotation curve.
In Sec. \ref{aquallike}, I describe briefly the special case of theories in which the gravitational Lagrangian is fashioned after AQUAL. Section \ref{usessec} lists some lessons that we can learn from such theories.
\section{Generalities   \label{generalities}}
Before presenting the specific families of mixed-MI--mixed-MG model theories, and discussing their properties, I bring up here several relevant points.

I said in the introduction that modifying inertia would logically imply having to modify gravity as well, because the gravitational action encapsulates the ``inertia of gravity'', such as would govern the propagation of gravitational waves, and cosmology. However, while this rational may apply for relativistic theories, here I deal with NR theories, in which the gravitational action is ``static'': It as no time dependence and it contains no time derivatives of the gravitational fields. As a result the field responds instantaneously to motion of the sources; it has no inertia. The above motivation does not then extend to the NR limit; in itself, it is not sufficient grounds for modifying the NR gravitational action. It is thus possible that modifying the relativistic gravitational action as part of generally modifying inertia, will not be reflected in the NR gravitational action, which will still be the Poisson action.
It is possible, however, that modifying the relativistic gravitational action, will lead also to a modified NR limit, even if this does not constitute a modification of the inertia of NR gravity.
\par
The particular theories I shall discuss below modify only the first (gravity) and last (inertia) terms in the Newtonian action (\ref{newtaction}), leaving the interaction term intact.
We note, however, that in taking the NR limit of general relativity, the second and last terms both fall from the particle action of general relativity, $-\int d\tau$. This brings to mind the possibility that if we modify inertia (i.e., the particle action) at the relativistic level, this will lead to a modification of both the second and third term of the NR action. On the other hand,
it is possible, as a realization Mach's principle, that inertia is due to interaction of the various DoFs with some omnipresent medium that impedes changes in the ``state of motion''. In such a scenario, our sought-for theory would be an effective result of eliminating the DoFs of that medium. The medium is then the primary giver, or  modifier of inertia.\footnote{It then must have its own basic inertia. There are many examples of such instances in physics where effective inertia emerges or is greatly modified; see discussions, e.g., in Refs. \cite{milgrom99,milgrom23}.} In this case it may be that only kinetic actions are modified, and are described as in MOND, not the interaction term.
\par
Be that as it may, not having in mind any specific such theory, we consider MOND extensions for each NR term separately.
At this level, there is an advantage in keeping the interaction term intact, in that this makes the theory invariant to constant shifts of the potential.
The retention of the Newtonian interaction has an additional advantage, at least in a heuristic context. The fact that it does not involve additional DoFs, or derivatives of $\f$, implies that the physical particle trajectories are governed solely by $\gf$. This will enable us to arrive at some useful predictions beyond what can be said for the more general case. Inasmuch as the modified gravitational Lagrangian involves additional DoFs, they will be only auxiliary, and appear only in the field equation for $\f$.

\section{The model theories  \label{theories}}
The most general mixed-MG-MI actions I consider here are of the form
\beq  I=\frac{\az^2}{G} \int_{t_1}^{t_2} \frac{dt}{T}~\Lf[\f,\Psi_k;\az] -\int_{t_1}^{t_2} \frac{dt}{T}~\sum_i m_i\f(\vvr_i(t))+\sum_i m_i\IK[\{\vvr_i(t)\};\az],   \eeqno{genaction}
where $T=t_2-t_1$, and the reason of the unconventional $T^{-1}$ factor in the first two action terms will be explained below.
\par
The gravitational Lagrangian, $\Lf$, is a functional of the gravitational potential, $\f$, and, possibly, of additional gravitational DoFs, $\Psi_k$, which can have various tensorial ranks. It is also a function of $\az$. $\Lf$ is possibly space-nonlocal; indeed, Refs. \cite{soussa03,deffayet14,woodard18,tan18,woodard24,deffayet26}
have proposed and discussed extensively relativistic, metric MOND theories that are nonlocal, with applications to galactic systems and to cosmology. The NR limits of such theories may result in NR gravitational actions that are space-nonlocal functionals of the potential.
We are dealing with NR theories, where the instantaneous values of the gravitational fields are determined by the instantaneous mass distribution. So $\Lf$ does not depend explicitly on time, and does not involve time derivatives of $\f$ and $\Psi_k$.
\par
One can study special cases of such a Lagrangian. For example, Lagrangian families constructed after AQUAL are briefly described in Sec. \ref{aquallike}, and a family constructed after QUMOND are discussed briefly in Appendix \ref{qumondapp}. But for now I keep the form of $\Lf$ as general as possible.

\par
In the particle kinetic action, $\IK$ is a universal functional of the particle trajectory $\{\vvr(t)\}$, and a function of $\az$. Its universality ensures the universality of free fall of test particles (the same trajectories are shared by particles of all masses). The universality of free fall of composite bodies has to be checked and ensured for candidate theories.
\par
The reason for the unconventional factors of $T^{-1}$ in the action is as follows: As explained in Ref. \cite{milgrom94}, the kinetic action may well be time nonlocal. Namely, it cannot be written as an integral over a finite time interval over some time-local Lagrangian.
Instead, it can be a functional of the full trajectory of the particle at all times.
In such a case, the expression for the action, Eq. (\ref{genaction}) only makes sense in the limit $t_1\rar -\infty$, $t_2\rar\infty$. The interaction term is time-local, and I assume that so is the gravitational Lagrangian $\Lf$. Still, to match the last term, we also write the first two action terms, formally, as functionals of the full time history of the systems, and I write them as time averages of the Lagrangians, so that if the Lagrangians themselves are bounded, the action remains finite in the limit. If $\IK$ is local, and underlain by a Lagrangian $L_K$, then $\IK=T^{-1}\int_{t_1}^{t_2}dt~L_K$; it is the time average of the kinetic Lagrangian for a unit mass.
\par
The gravitational Lagrangian, $\Lf$, has the general form of the gravitational Lagrangians considered in Ref. \cite{milgrom14b}, and the kinetic action is of the general form that was discussed extensively in Ref. \cite{milgrom94}. Here I draw much from results in these papers. However, Refs. \cite{milgrom14b,milgrom94} dealt, respectively, with pure-MG, and pure-MI MOND theories, hence the respective action terms had MOND-specific properties -- with specific implications that we cannot employ here. Here, in general, neither $\Lf$, nor $\IK$, in themselves, have MOND properties. If either of them is paired with a Newtonian version of the other we do not get a MOND theory. It is only their combined effects that give the required MOND behaviour (e.g., the correct DML). They are, however, assumed to each have the respective Newtonian limits, in the high acceleration limit, $\az\rar 0$, as they appear in Eq. (\ref{newtaction}).
\par
We shall use dimensional arguments extensively; so we need to establish the dimensions of the different quantities. The potential $\f$ and the kinetic $\IK$ have dimensions of squared velocity. Without loss of generality, we can assume that $\Psi_k$ too have no mass dimensions (otherwise we can renormalize them by powers of $G$). The $\az^2$ prefactor is added for convenience, giving $\Lf$ dimensions $[\ell]^3$.
\par
We deal with gravitational actions where the potentials enter the Lagrangian locally in time, and with no time derivative. So,
extremizing the action over variations of $\f$ and $\Psi_k$ amounts to doing it for a fixed time. Write for variations $\d\f(\vvr),~\d\Psi_k(\vvr)$,
\beq  \d\Lf=\int~\drt[X_{\f}\d\f(\vvr)+\sum_k X_{\Psi_k}\d\Psi_k(\vvr)]~~~+~~{\rm S.T.}, \eeqno{mayure}
and extremizing the action under such variations that annihilate the surface term ${\rm S.T.}$, we get the field equations.
 \beq X_{\Psi_k}\equiv\frac{\d\Lf}{\d\Psi_k}=0,~~~~~ X_{\f}\equiv\frac{\d\Lf}{\d\f}=\az^{-2}G\r.  \eeqno{manud}
\par
 To get the particle equation of motion, we write the variation in the kinetic action (dropping the particle index) under variations $\vvr(t)\rar\vvr(t)+\d\vvr(t)$ as
\beq  \d\IK=-\int_{t_1}^{t_2} \frac{dt}{T}\A[\{\vvr(t)\};t,\az]\cdot\d\vvr(t)+
\frac{1}{T}[\eff\cdot\d\vvr(t)]\^{t_2}\_{t_1}, \eeqno{mateq}
where $\A$ has dimensions of acceleration, and is a functional of $\{\vvr(t)\}$.
The second term on the right-hand side is the difference between the values of $\eff\cdot\d\vvr(t)$, at the two end times; $\eff$ is some linear vector operator, acting on functions of $t$, [and may depend on $\vvr(t)$].\footnote{The role of $\eff$ in defining conserved quantities was discussed in Ref. \cite{milgrom94}.} For nonlocal kinetic actions, the limit $t_1\rar -\infty$, $t_2\rar\infty$ is understood.
The particle equation of motion is gotten, as usual, by requiring that physical trajectories extremize the action under arbitrary $\d\vvr(t)$ that annihilate the end term. This gives
\beq  \A[\{\vvr(t)\};t,\az]=-\gf(\vvr(t)).  \eeqno{muriso}
\par
We shall restrict ourselves to theories that are invariant to constant shifts in $\f$. We see from Eq. (\ref{muriso}) that particle dynamics is dictated by $\gf$, which is so invariant. And so $\Lf$ is required to change by a constant under such shifts, which thus do not affect the field equations.
\par
As in pure MG theories, the concept of ``phantom matter'', $\r_p$, may be useful in describing the solution of the gravitational field. It is defined such that $\r_p=(4\pi G)^{-1}\Delta\f-\r$. But it is not the MOND proxy for dark matter, since the particle acceleration is not $-\gf$ , as in ND.

 \subsection{Virial relations}
Recapping a derivation in Ref. \cite{milgrom94}, we note that since $\IK$ has dimensions of $[\ell]^2[t]^{-2}$, invariance to rescaling the units of length implies that
\beq   \IK[\{\l\vvr_i(t)\};\l\az]=\l^2\IK[\{\vvr_i(t)\};\az].   \eeqno{jayretw}
Taking $\l=1+\epsilon$, and taking the first order in the infinitesimal $\epsilon$, this implies, using Eq. (\ref{mateq}),
 \beq   2\IK=\az\frac{\partial\IK}{\partial \az}-\int_{t_1}^{t_2} \frac{dt}{T}\A[\{\vvr(t)\};t,\az]\cdot\vvr(t)+
\frac{1}{T}[\eff\cdot\vvr(t)]\^{t_2}\_{t_1}. \eeqno{matsua}
\par
Virial relations apply to trajectories solve the EoM, and for which the difference in end terms remains finite,\footnote{In deriving the equation of motion we required stationarity of the action to $\d\vvr(t)$ that vanish at the end points. Here, $\d\vvr(t)=\epsilon\vvr(t)$ and does not vanish at the end points.} for example, if the trajectory is periodic, or is, more generally, confined to a finite volume. So, taking $T\rar\infty$, and using the equation of motion, we get
 \beq   \IK\left(2-\frac{\partial\ln{\IK}}{\partial \ln\az}\right)=\avt{\vvr(t)\cdot\gf}, \eeqno{matsmio}
 where $\langle\rangle_t$ stands for the long-time average, and $\IK$ is calculated for the full trajectory (for time-local theories it is the long-time average of the kinetic Lagrangian).
This relation holds for each particle separately. However, in most applications we can only measure a one-time snapshot of the N-body system under study. So, to get a useful relation, we should take some average over the system. Such a system average can be almost time independent for quasi-stationary systems that vary globally over times much longer than the dynamical times over which the individual-particle attributes change appreciably. I shall assume this, and drop the time average. For example we could take the mass-weighted average, and designate it $\langle\rangle_m$:
 \beq   \langle\IK\rangle_m\left(2-\frac{\partial\ln\langle\IK\rangle_m}{\partial \ln\az}\right)=\avm{\vvr\cdot\gf}. \eeqno{matsgut}
In the high-acceleration limit the above virial relations reduce to the Newtonian ones; e.g., the single-particle relation (\ref{matsmio}) reduces to $\avt{\dot\vvr^2}=-\avt{\fN}$.

\par
Another, potentially useful, relation can be derived as follows: Considering a change of the units of time\footnote{Here we cannot use the argument based on scaling the length units, because there may appear space derivatives in the arguments of $\Lf$.} by a factor $\l$, and note that $\Lf$ has no time dimensions, and that neither time, nor time derivatives, appear in its arguments, we have
\beq \Lf[\l^2\f,\l^{-\a_k}\Psi_k;\l^2\az]=\Lf[\f,\Psi_k;\az]  \eeqno{matateq}
where $\a_k$ is the time dimension of $\Psi_k$. Taking $\l=1+\epsilon$, and taking the first order in the infinitesimal $\epsilon$, this implies, using Eq. (\ref{mayure}),
\beq  \int~\drt[2X_{\f}\f(\vvr)-\sum_k \a_k X_{\Psi_k}\Psi_k(\vvr)]+2\az\frac{\partial\Lf}{\partial\az}~~~+~~{\rm S.T.}=0  \eeqno{kurwei}
Applying this relation to $\f$ and $\Psi_k$ that solve the field equations, and that also annihilate the surface terms,\footnote{This condition has to be checked for the particular theory under study. For example, for AQUAL as pure MG, it is not satisfied. But, all the other theories in the family I describe in Sec. \ref{aquallike} of which AQUAL is an end point, do satisfy this condition.} and using the field equations (\ref{manud}), we get
\beq \av{\f}\equiv \frac{1}{M}\int\drt~\r\f=-\frac{\az^3}{MG}\frac{\partial\Lf}{\partial\az}\Big|_{\f,\Psi_k}.    \eeqno{garemop}
In the Newtonian limit $\Lf=-(8\pi \az^2)^{-1}\int\drt~(\gfN)^2$, which gives $\av{\f}=-(4\pi MG)^{-1}\int\drt(\gfN)^2$, which, otherwise, follows straightforwardly from the Poisson equation.

 \section{The deep-MOND limit and scale invariance \label{dmlsec}}
The defining requirement of the DML is that the theory becomes spacetime scale invariant (SI) in this limit \cite{milgrom09}, from which follows some of the most salient, primary predictions of MOND (see, e.g., Refs. \cite{milgrom09,milgrom14a,milgrom25}).
\par
Thus, the DML equations of motion of the MOND theories underlain by the action (\ref{action}), has to be invariant under
\beq   (t,\vvr)\rar\l(t,\vvr).
    \eeqno{scalala}
Under such transformations, quantities such as times and radii, as well as time derivatives of $\vvr(t)$, such as velocities and accelerations, transform according to their $[m][\ell][t]$ dimensions; e.g., $\vv(t)\rar\vv(t/\l)$ and $\va(t)\rar\l^{-1}\va(t/\l)$. But other quantities, such as $\f$ and $\Psi_k$ in our case here, do not, generally have well-defined scaling behavior. For example, in pure-MG theories, $\f$ behaves as $r^{-1}$ in the high-acceleration regime, changing gradually to $\ln{r}$ in the DML; it doesn't have well-defined scaling properties.
However, we require that in the DML such quantities are multiplied  by a power of $\l$ under the scaling of Eq. (\ref{scalala}); the power being their DML scaling dimension (SD).
\par
These quantities
may be assigned a SD that does not correspond to their $[m][\ell][t]$ dimensions, if by so assigning we end up with a theory that is SI. For example, if $\Psi$ has dimensions $[m]^a[\ell]^b[t]^c$, it could still be dictated that it transforms under scaling as $\Psi\rar\l^{-\chi}\Psi$ with $\chi\not =-b-c$.
\par
It is sometimes convenient to normalize all the degrees of freedom into ``standardized'' ones (by multiplying them by the appropriate power of $\az$), so that their new $[m][\ell][t]$ dimensions do match their assigned SD. With such normalization of the DoFs, SI implies that only the combination $\azg\equiv\az G$ can appear in DML expression, neither $\az$ nor $G$ separately \cite{milgrom09,milgrom14a,milgrom25}.\footnote{Time and coordinates that enter the action through derivatives already have these properties, so their appearance does not stand in the way of our argument.}
The sole appearance of the dimensionful constant $\azg$ then implies a two-parameter scaling invariance of the DML under
\beq m\rar \a\_m m,~~~~\vvr\rar \a\_l\vvr,~~~~t\rar \a\_t t , ~~~(\a\_t\equiv \a\_l\a\_m^{-1/4}),  \eeqno{sculam}
which, when viewed as changes of units do not change the value of $\azg$. In terms of the system's DoFs these read
\beq \vvr(t)\rar \a\_l\vvr(t/\a\_t),~~~~\vv(t)\rar \a\_m^{1/4}\vv(t/\a\_t)~~~~
 \va(t)\rar \a\_m^{1/2}\a\_l^{-1}\va(t/\a\_t),~~~~\vgN(t)\rar \a\_m\a\_l^{-2}\vgN(t/\a\_t).  \eeqno{sculap}

\par
These scaling properties of the DML do not hinge on using the standardized versions of the auxiliary DoFs; this choice is only convenient for deriving the scaling laws. In the end those laws relate only to masses, times, distances, velocities, accelerations, etc. which do not require standardization.

\subsection{Form of the DML action   \label{dmlactionsec}}
To ensure SI of the DML of the theory, the DML limit of the action,  $I\^D$, has to have a well defined SD; so that under scaling $I^D\rar\l^{-\eta}I^D$, for some $\eta$, and all terms in the action scale in this way. This results in the equations of motion themselves being SI. The form of second term in the action (\ref{genaction}) then implies that $\f$ has SD $-\eta$: $\f(\vvr)\rar\l^{-\eta}\f(\vvr/\l)$. Since its dimensions are $[\f]=[\ell]^2[t]^{-2}$, which do not match its SD of $-\eta$, except for $\eta=0$, we define its standardized representation as $\tilde\f=\az^\eta\f$.
\par
The DML of $\IK$ has to have the SD $-\eta$, as well. Using dimensional argument we can see that
\beq   \IK\rar\IK\^D= \az^{-\eta}\U[\{\vvr(t)\}],    \eeqno{xvii}
where $\U$ is some universal functional of the trajectory alone, with no dependence on dimensionful constants. To deduce this dependence of $\IK\^D$ on $\az$, consider the following: Let the trajectory $\vvr(t)$ be scaled to the trajectory $\hat\vvr(\hat t)=\l\vvr(\hat t/\l)$. The value of the action for the scaled trajectory then has to be $\IK\^D[\{\hat\vvr(\hat t)\};\az]=\l^{-\eta}\IK\^D[\{\vvr(t)\};\az]$. $\IK\^D$ has dimensions of $[\ell]^2[t]^{-2}$. Thus, the value of $\IK\^D$ is unchanged if we change the units of length and time, multiplying them by the same factor $\l$. In the new unites the trajectory $\hat\vvr(\hat t)$ and the time as it appears in its argument attain back the values as in $\vvr(t)$, while the constant $\az$ has the new value $\l\az$ (instead of the value $\az$ in the old units). We thus have, $\IK\^D[\{\vvr(t)\};\l\az]=\IK\^D[\{\hat\vvr(\hat t)\};\az]=\l^{-\eta}\IK\^D[\{\vvr(t)\};\az]$. Taking $\l=\az^{-1}$ then gives $\IK\^D[\{\vvr(t)\};\az]=\az^{-\eta}\IK\^D[\{\vvr(t)\};1]\equiv\az^{-\eta}\U[\{\vvr(t)\}].$
\par
Similar arguments for $\Lf$ -- whose dimensions are $[\ell]^3$ -- show that its DML dependence on $\az$ is given by
\beq \Lf[\f,\Psi_k;\az]\rar \Lf\^D[\f,\Psi_k;\az]=\az^{-(\eta+3)}\N[\tilde\f,\tilde\Psi_k]=\az^{-(\eta+3)}\N[\az^\eta\f,\az^{\b_k}\Psi_k],   \eeqno{marediq}
where, $\tilde\f$ and $\tilde\Psi_k$ are standardized forms of $\f$ and $\Psi_k$, $\N[\tilde\f,\tilde\Psi_k]$ does not depend on dimensionful constants, and $\b_k$ are the powers of $\az$ needed to convert $\Psi_k$ to $\tilde\Psi_k$ (which we do not have to know for what follows).\footnote{Note that if $\N$ is written as a functional of non-standardized DoFs, $G$ and $\az$ can appear in its functional form.}
The DML action itself has dimensions $[I\^D]=[m][\ell]^2[t]^{-2}$; so, its standardized form is $\tilde I\^D=\az^\eta I$. Writing the DML action in terms of standardized quantities
we have
\beq \tilde I\^D=\frac{1}{\azg} \int_{t_1}^{t_2} \frac{dt}{T}~\N[\tilde\f,\tilde\Psi_k] -\int_{t_1}^{t_2} \frac{dt}{T}~\sum_i m_i\tilde\f(\vvr_i(t))+\sum_i m_i\U[\{\vvr_i(t)\}].   \eeqno{dmlstanaction}
As expected, $\azg$ is the only constant that appears, and with a simple dependence.
\par
The reverse also holds: all theories with an action of the form (\ref{dmlstanaction}) are SI: Because all the DoFs are standardized, scaling has the same effect on this expression as changing the units of length and time by the same factor. But this does not change the value of $\azg$; thus, scaling multiplies such an action by an overall factor that depends on its dimensions. The resulting equations of motion are thus SI.
\par
It is straightforward to ascertain directly that in a theory governed by an action of the form (\ref{dmlstanaction}), the two-parameter scaling of Eq. (\ref{sculap}) are symmetries, in the sense that applying them to any solution of the theory yields another solution.

\subsection{DML virial relations}
In Ref. \cite{milgrom14b} a DML virial relation was derived in the class of theories that correspond to $\eta=0$ here (with the proviso that the kinetic action is Newtonian). This relation reads
\beq \avm{V^2}=\frac{2}{3}(M\azg)^{1/2}\left[1-\sum_i \left(\frac{m_i}{M}\right)^{3/2}\right].   \eeqno{vivared}
How does this generalize to the theories with $\eta\not = 0$?
\par
Since the dependence of $\IK$ on $\az$ in the DML is known from Eq. (\ref{xvii}), the general virial relations Eq. (\ref{matsgut}) reduces to
 \beq   (2+\eta)\avm{\IK\^D}=\avm{\vvr\cdot\gf}. \eeqno{matsjan}
In terms of the standardized quantities we have
 \beq   (2+\eta)\avm{\U}=\avm{\vvr\cdot\grad\tilde\f}. \eeqno{matshul}
To specialize the general field virial relation (\ref{garemop}) to the DML, we need the DML dependence of $\Lf$ on $\az$ when the Lagrangian is written in terms of the nonstandardized DoFs. The last equality in Eq. (\ref{marediq}) is sufficient for this purpose. But it is simpler to
start from the standardized form of the DML action Eq.(\ref{dmlstanaction}), noting that the time dimension of $\tilde\f$ is $-2(\eta+1)$, and that of $\N$ is $-2(\eta+3)$, and
repeating the arguments that led to the field virial relation (\ref{garemop}).\footnote{We use the fact that on dimensional grounds (the effect of changing the unit of time by a factor $\l$) $\N[\l^{2(\eta+1)}\tilde\f,\l\^{2\a_k}\tilde\Psi_k]=\l^{2(\eta+3)}\N$. Then take the $\l$ derivative at $\l=1$, and use the field equation.} We get (for solutions of the DML equations that annihilate the surface terms)
\beq  \frac{\eta+3}{\eta+1}\azg^{-1}\N[\tilde\f,\tilde\Psi_k]=\int\drt~\r\tilde\f,           \eeqno{laterwqa}
which can be written in terms of $\Lf[\f,\Psi_k,\az]$ as
\beq  \frac{\eta+3}{\eta+1}\frac{\az^2}{MG}\Lf=\av{\f}.           \eeqno{latermaz}
 \par
The virial relation (\ref{matshul}) is exact in the DML.\footnote{Relation (\ref{laterwqa}) [or (\ref{latermaz})] is only useful for studies of field properties, but is not directly applicable to observed dynamics as it does not refer to the particle DoFs.} However, in practice, we cannot employ it as it is written, because it is impractical to measure the quantities that enter this relation. Instead -- as is done when employing the Newtonian virial relation -- we approximate the two sides of the relation by some expressions that involve characteristic velocity, $\sigma$, and radius, $R$, of a system, and its total mass $M$. But which expressions built from these attributes can we use? We cannot decide solely based on the dimensions, since we can construct various expressions of the same dimensions with $\azg$, $\sigma$, $R$, and $M$, given that $\s^4/M\azg$ is dimensionless.
The expressions we form also have to have the same scaling properties as the quantities they are to represent. For example, $\avm{\U}$ is an average over quantities that depend only on the trajectories. If we scale all masses by the same factor it does not change; so, it cannot have $M$ appearing. Its dimensions and scaling dimensions are those of $\sigma^{2(\eta+1)}/R^{\eta}$, and it can be written only as such an expression, with a dimensionless prefactor.
\par
On the other hand, $\tilde\f$ is oblivious to velocities and depends only on masses and their positions. The only expression that one can write involving $M$, $R$, and possibly $\azg$ with the dimensions and scaling dimension of $\avm{\vvr\cdot\grad\tilde\f}$ is $(M\azg)^{(\eta+1)/2}/R^\eta$, again with some dimensionless prefactor. Equating the two, we get the DML $M-\sigma$ relation
\beq  \frac{\sigma^4}{M\azg}=\xi,    \eeqno{lanawer}
where $\xi$ is a dimensionless number that can depend on the theory, on the exact definitions of $R$ and $\sigma$, and on dimensionless attributes of the system under study, such as geometrical aspect ratios or various mass ratios.\footnote{Because $\avm{U}$ and $\avm{\vvr\cdot\grad\tilde\f}$ scale in the same way with size, $R$ drops from the equality. However, the prefactors in the two expressions may depend on the exact definition of $R$.}
\par
In a given theory, $\xi$ takes the same value for all DML systems, of all sizes, and all masses, characterized by the same values of dimensionless attributes; for example, for all DML exponential discs with the same ratio of scale height to scale length. This holds no matter how we define sigma, as long as it is defined in terms of dimensionless attributes,\footnote{For example, for exponential discs it may be arbitrarily taken as the rotational speed in the midplane at 3.5 scale lengths.} and it follows from only the scaling properties of the DML, Eq. (\ref{sculap}). However, with a proper choice of $\s$, it follows from the basic tenets that $\xi$ has a value of order unity for all DML systems, making relation (\ref{lanawer}) a predicted universal DML that can be considered tight since it can be applied to galactic systems with a span of masses of many orders of magnitude \cite{milgrom19,yong26}.
\par
One of the requirements that underlie a MOND theory is that it transits from the DML to the Newtonian regime within a range of accelerations of order $\az$. This follows from the requirement that the theory does not introduce dimensionless parameters that differ greatly from unity (so $\az$ defines both the location and the width of the transition).
It was shown in Ref. \cite{milgrom25} that in such a MOND theory, with a proper definitions of $\sigma$ and $R$, indeed, $\xi$ has to be of order unity. This is essentially because the general virial relation Eq. (\ref{matsgut}) then transits from the Newtonian relation $\s^2\approx MG/R$ to the DML one, Eq. (\ref{lanawer}), and at the transition, where $\s^2/R\approx\az$, both the newtonian and the DML relations have to hold approximately,\footnote{This is where the proper definition of $\s$ has to be made. It has to be such that with an appropriate choice of some $R$, $\s^2/R$ is a good measure of whether the system is in the MOND or Newtonian regimes, and such that when we are in the Newtonian regime, $\s$, $R$ and $M$ satisfy the Newtonian virial relation.} which leads to $\xi\approx 1$.
\par
We note that the case $\eta=0$, which corresponds to pure MG, is special in that $R$ does not appear on either side of the virial relation (\ref{matsjan}) [or (\ref{matshul})]. The reason is that in this case each term in the action is SI, whereas for $\eta\not =0$, we only have that all terms scale in the same way, while not being individually SI. These large class of $\eta=0$ theories, which are the subject of Ref. \cite{milgrom14b}, are also special in that their DML relation (\ref{vivared}) is universal if $\s$ is defined as the mass-weighted root-mean-squared velocity, which follows directly from the DML scalings of Eq. (\ref{sculap}), without having to resort to what happens at the DML-Newtonian transition.
\par
Indeed, generally, pure-MG theories have $\eta=0$; seen by comparing the last two terms in the action, and noting that in this case the kinetic action is Newtonian and is invariant to scaling. At the other end, the DML of a pure-MI theories has $\eta=1$, because it is governed by the Poisson Lagrangian; so, comparing the first two terms in the action we see that $\f$ has to scale like $\int\drt(\gf)^2$ leading to $\eta=1$.\footnote{This clearly follows, more directly, from the fact that in this case $\f$ is the Newtonian potential and scales as $r^{-1}$.}
Is the inverse also true? For $\eta=0$, we see from Eq. (\ref{xvii}), that the DML kinetic action cannot depend on $\az$. Barring some peculiarity, this points to the kinetic action being the Newtonian one, namely to a pure-MG gravity. But it is not clear from expression (\ref{marediq}) of the DML field Lagrangian that for $\eta=1$, $\az$ disappears from this expression. So, it remains to be checked if there are acceptable theories with $\eta=1$ that do modify gravity. We do note, however, that for $\eta=1$ the potential scales like $r^{-1}$, as in the Newtonian case. In particular, we see from Eq. (\ref{xiv}) below for the asymptotic potential, that asymptotically $\f\propto MG/r$. So, $\f$ must at least share these properties with the Newtonian potential.
\par
The range $0\le \eta\le 1$ thus span a range of theories from pure MG to pure MI.
An interesting question, which I shall not try to answer here is whether there are acceptable MOND  theories with $\eta<0$ and $\eta>1$.
\section{Additional aspects of the theories  \label{rotasym}}
We can make some additional inroads for more specific subclasses of theories, or for specific phenomenologies in the general case. For example, in Sec. \ref{aquallike}, I discusses further simplifications for a subclass of theories where the gravitational Lagrangian is space local and generalizes the AQUAL Lagrangian to the case $\eta\not =0$. Here I discuss some additional relevant aspects of the more general class of theories, such as the asymptotic behavior of the potential, and rotation curves.
\subsection{Asymptotic behavior of the potential   \label{asympot}}
One of the basic requirements of MOND is that far from an {\it isolated} bounded body of total mass $M$, test-particle dynamics depend only on $M$, and not on other attributes of the body \cite{milgrom14b,milgrom25}. If indeed the gravitational sector of our theories has this property, then, it should have a unique solution for $\f$ for an ideal, isolated, central point mass. Since the only dimensionless expression one can form from $M$, $G$, $\az$, and $r$ must be a function of $\rM/r=(MG/\az)^{1/2}/r$, where $\rM$ is the MOND radius of the mass $M$, this solution must be of the form
\beq \f\_{pm}=(M\azg)^{1/2}f\left(\frac{\rM}{r}\right).  \eeqno{moretq}
The gradient of $\f$ -- which is equated with $|\A|$ in the EoM -- can then be written as
\beq  g\_{pm}=\Big|\frac{d\f\_{pm}}{dr}\Big|=\gN\n\_{pm}\left(\frac{\gN}{\az}\right),  \eeqno{xivre}
where $\n\_{pm}$ is the theory's interpolating function relevant for the case of a point mass.
To restore the Newtonian limit for $\gN\gg\az$ we need $\n\_{pm}(y\gg 1)\approx 1$. In the opposite limit $\f$ has to attain a SD $-\eta$; so, for $\eta\not =0$, it has to behave asymptotically as
\beq  \f\rar\f\_{pm}\^{as}= -(\gamma/\eta)(M\azg)^{1/2}\left(\frac{\rM}{r}\right)^{\eta},  \eeqno{xiv}
(with $\gamma$ a number),\footnote{The factor $1/\eta$ in front of expression (\ref{xiv}) ensures that $\gamma$ remains finite in the limit $\eta\rar 0$, which should give a legitimate theory. In this limit, $-\eta^{-1}r^{-\eta}\rar-\eta^{-1}+\ln{r}$, and we consider theories that are invariant to constant shifts in $\f$. Indeed, the gradient of the former goes to that of the latter in the limit, as seen in Eq. (\ref{xivas}).}
and
\beq  g\_{pm}\^{as}=\gamma(M\azg)^{1/2}\frac{\rM^\eta}{r^{1+\eta}}= \gamma\az\left(\frac{\gN}{\az}\right)^{\frac{1+\eta}{2}} \eeqno{xivas}
[$\n\_{pm}(y)\propto y^{(\eta-1)/2}$]. For MG, where $\eta=0$, the asymptotic behavior of $\f$ is logarithmic, $\f\_{as}= \gamma(M\azg)^{1/2}\ln{r}$, up to an additive constant.
\par
Expression (\ref{xiv}) has $G$ and $\az$ appearing separately (in the expression for $\rM$), because $\f$ is not standardized.
For the standardized potential $\tilde\f$ we have
\beq  \tilde\f\_{pm}\^{as}= -(\gamma/\eta)(M\azg)^{(1+\eta)/2}r^{-\eta}.  \eeqno{xviii}
\par
In general, there are additional terms in the asymptotic behavior (analogous to higher multipoles in ND), but they are expected to be of higher order in $\ell/r$ where $\ell$ is the characteristic size of the central body.\footnote{For example, the multipole expansion in AQUAL is described in Ref. \cite{milgrom86}.}
\par
As regards a finite-size central mass, our basic assumption implies that, far enough from the central mass, expression (\ref{xivre}) is a good approximation, whether its extent is smaller or larger than its MOND radius. So, in any event, the asymptotic DML expressions (\ref{xiv})(\ref{xivas}) are valid.
\par
An interesting possible exception to the above asymptotic behavior can appear in the context of the NR limit of some bimetric MOND theories (BIMOND) with a ``twin matter'' sector \cite{milgrom10b}, where in the DML the twin matter acts on matter as a negative gravitational mass would. In these theories, the DML field equation for the matter potential is an AQUAL-like equation sourced by the difference between the matter and twin-matter densities.\footnote{Matter and twin matter self attract as usual, but they repel each other, and they have positive inertial masses.} If our gravitational Lagrangian is constructed after the fashion of such a theory, for a system made of matter and twin matter of equal masses, the asymptotic potential felt by matter would correspond to zero central mass -- so the asymptotic term that depends on the total mass, as in Eq. (\ref{xiv}), is absent -- and $\f\^{as}$ would depend on the distribution of the masses within the central body. For example, in the pure-MG case ($\eta=0$), the DML gravity is conformally invariant \cite{milgrom97}. This can be used to solve exactly the potential for two matter and twin-matter point masses $M$, a distance $\ell$ apart,
which behaves asymptotically as $\f\propto (M\azg)^{1/2}\ell z/r^2$ ($z$ is the coordinate along the dipole axis) \cite{milgrom97}.

\subsection{Rotation curves   \label{rotcurkin}}
The kinetic side of the EoM is much simplified when we come to describe exactly circular trajectories (of constant speed) in the midplane of an axisymmetric and plane-symmetric mass distribution. Though limited, these are still important circumstances, since such orbits are an idealized description of the orbits of test particles (stars and gas) from which rotation curves of disc galaxies are measured.\footnote{In reality, these orbits are not exactly circular, in that they do wonder somewhat off the symmetry plane, and of the circular orbit.}
As has been argued in Refs. \cite{milgrom94,milgrom22a,milgrom23}, such ideal orbits are completely specified by their radius, $r$, and orbital speed, $V$. Thus, the functional $\A$ in Eq. (\ref{muriso}) becomes {\it a function} of $r$ and $V$. The assumed symmetries of the kinetic action (rotational, reflection and time reversal) imply that $\A$
for circular orbits is parallel to $\vvr$. So, since $\A$ has dimensions of acceleration, we can write it, in all generality, as
\beq   \A=-\frac{V^2}{r}\m_k\left(\frac{V^2}{r\az}\right)\frac{\vvr}{r},  \eeqno{fepoleq}
where $\m_k$ is universal (for a given theory). The EoM can thus be written as a functional relation between the MOND acceleration $a=V^2/r$ and the auxiliary acceleration $g\_\f=|\gf|$
\beq  a\m_k(a/\az)=g\_\f  ~~~~~{\rm or}  ~~~~~ a=g\_\f\n_k(g\_\f/\az).    \eeqno{balamala}
\par
The ``kinetic interpolation functions'', $\m_k$ and $\n_k$, are only relevant for ideal circular trajectories, and are examples of interpolating functions that emerge in MOND theories, but that are relevant for only a limited set of phenomena.
\par
As was shown in Ref. \cite{milgrom94}, $\m_k$ can be expressed directly in terms of the action, without having to derive the field equation that defines $\A$. Here, it follows by substituting relation (\ref{fepoleq}) in Eq. (\ref{matsua}) to get
 \beq  V^2\m_k\left(\frac{V^2}{r\az}\right)= \IK^c\left(2-\frac{\partial\ln\IK^c}{\partial \ln\az}\right), \eeqno{matreca}
 where $\IK^c$ is the kinetic action reduced to its form for circular orbits. It has dimensions of squared velocity, and can depend only on $V$, $r$, and $\az$, so must be of the form
 \beq  \IK^c=V^2\J\left(\frac{V^2}{r\az}\right). \eeqno{pilopil}
 Thus,
 \beq  \m_k(x)= \J(x)\left(2+\frac{\partial\ln\J}{\partial \ln{x}}\right). \eeqno{matumna}
This expression can be a very useful shortcut for calculating $\m_k$, since the general kinetic action can be a rather complicated functional of the trajectory, while is expression for circular orbits may be much simpler.
\par
Unlike pure MI, here we do not get an algebraic relation between the MOND acceleration, $a$ and $\gN$ on circular trajectories. While $a$ is a function of $|\gf|$, the latter is not a function of the Newtonian acceleration. But even so, the above results show that in such theories one is left with solving only the gravitational equations to obtain $\f$, while the possibly hard problem of solving nonlocal equations of the kinematic side becomes much easier.
\par
For a pure-MI theory, $\m_k$ has the standard limiting behaviors. For mixed theories it does not. While it is required to tend to unity for high accelerations, where $\A$ becomes the kinetic acceleration, its DML behavior depends on $\eta$.
In the limit of small arguments, since $\IK$ has SD $-\eta$, so does $\J$, and hence $\J(x)\propto x^\eta$, which gives from Eq. (\ref{matumna})
\beq \m_k\rar \m\^D_k(x)=\zeta x^\eta, ~~~~~~ \n_k\rar \n\^D_k(y)=\zeta^{\frac{-1}{1+\eta}} y^{\frac{-\eta}{1+\eta}},\eeqno{ghagha}
where $\z$ is a number.
\par
This lets us express the DML kinetic action for circular orbits as a function of $V$ and $r$:
\beq \U^c(V,r)=\frac{\z}{2+\eta}\frac{V^{2(1+\eta)}}{r^\eta}.   \eeqno{kiutfed}

In particular, for the asymptotic rotation curve, we can combine expression (\ref{ghagha}) with the behavior of $\f'$ in Eq. (\ref{xiv}), to get
\beq V(r)=V\_\infty=\left(\frac{\gamma}{\z}\right)^{\frac{1}{2(1+\eta)}}(M\azg)^{1/4}, \eeqno{asympv}
namely, a constant asymptotic rotational speed, and the mass-asymptotic-speed relation (MASR) between $M$ and the asymptotic speed. These two results are expected from the outset, as they follow inescapably from only SI and the sole appearance of $\az$ as a dimensionful constant besides $G$, which our theories satisfy.
\par
One standardly defines the normalization of $\az$ such that the universal MASR Eq. (\ref{asympv})
reads $V\_\infty=(M\azg)^{1/4}$. This can be arranged, noting that in their definitions, $\gamma$ and $\z$ appear multiplied by different powers of $\az$, i.e., $\gamma\az^{(1-\eta)/2}$ [in Eq. (\ref{xiv})] and $\z\az^{-\eta}$ [in Eq. (\ref{ghagha})]. We can thus take for  $\az$ a normalization such that $\gamma=\z$, and I assume that this is already the choice of $\az$ that I started with in Eq. (\ref{genaction}).
This means that the requirement $\gamma=\z$ is already built into the theory.
This procedure works as follows: Starting with a given theory characterized by some dimensionless parameters of order unity\footnote{A basic requirement from a MOND theory is that it does not involve dimensionless parameters much different from unity (see, e.g., \cite{milgrom25}).}; calculate $\gamma$ and $\z$ as functions of these parameters; then impose the constraint $\gamma=\z$ on these parameters, and proceed with the resulting theory. This automatically ensures the required normalization of $\az$ that appears in the theory.

\section{Examples: the MG action modelled after AQUAL  \label{aquallike}}
Among the possible gravitational Lagrangian, $\Lf$, we may consider local Lagrangians of the form
\beq  \Lf=\int \drt  \L_f[\f,\Psi_k;\az],  \eeqno{motibufa}
where $\L_f$ is a function of the gravitational DoFs and of their finite-order derivatives.\footnote{AQUAL, QUMOND, and TRIMOND are examples of such theories with $\eta=0$.}
\par
 Specifically, I exemplify some of the results above for a choice of $\L_f$ that is modeled after AQUAL \cite{bm84}, namely, where $\f$ is the only gravitational DoF, and where $\L_f$ is a function of only $\gf$ [rotational invariance that dictates that it is a function of $(\gf)^2$]. The case $\eta=0$ of this family is AQUAL itself.
\par
We thus proceed with the family of theories underlain by the actions of the form
\beq  I=-\frac{\az^2}{8\pi G} \int_{t_1}^{t_2} \frac{dt}{T}~\drt \hat\F\left[\frac{(\gf)^2}{\az^2}\right] -\sum_i m_i\int_{t_1}^{t_2} \frac{dt}{T}~\f(\vvr_i(t))+\sum_i m_i\IK[\{\vvr_i(t)\};\az],    \eeqno{action}
where $\hat\F$ is not the AQUAL $\F$, but is still required to have the Newtonian limit $\hat\F(u)/u\rar 1$, for $u\gg 1$.
\par
Variation over $\f$ gives a field equation that looks formally like that of AQUAL
\beq \div\left[\m\_G\left(\frac{|\gf|}{\az}\right)\gf\right]=4\pi G\r,  \eeqno{aqlik}
where $\m\_G(x)\equiv\hat\F'(x^2)$. This ``gravitational interpolating function'' is not related to the kinetic IF discussed in Sec. \ref{rotcurkin} in relation to circular orbits.

\subsection{Spherical systems}
For a one-dimensional (e.g., spherical) mass distribution, application of Gauss's theorem to Eq. (\ref{aqlik}) shows that $|\gf|$ is a function of $\gN=|\vgN|$ (the two accelerations are parallel and point towards the center):
\beq  \m\_G\left(\frac{|\gf|}{\az}\right)|\gf|=\gN, ~~~~~~{\rm or}~~~~~~|\gf|=\gN\n\_G\^{1d}\left(\frac{\gN}{\az}\right). \eeqno{x}
Note that $\n\_G\^{1d}$ is relevant only for one-dimensional configurations, and that it coincides with the point-mass interpolating function, $\n\_{pm}$ of the theory, as defined in Eq. (\ref{xivre}), since a point mass is a special case of a spherical system.
\par
As demonstrated in Ref. \cite{brada95}, the field equation (\ref{aqlik}) can also be solved analytically for disc-plus-bulge galaxy models constructed on the Kuzmin disc.

 For circular motions in a symmetry plane of such a system we have from the EoM and Eq. (\ref{balamala}) a functional relation between the MOND acceleration a and $\gN$
 \beq a\m_k\left(\frac{a}{\az}\right)= \gN\n\_G\^{1d}\left(\frac{\gN}{\az}\right),    \eeqno{nukvar}
 or explicitly
  \beq a= \gN\n\_G\^{1d}\left(\frac{\gN}{\az}\right)\n_k\left[\left(\frac{\gN}{\az}\right)\n\_G\^{1d}\left(\frac{\gN}{\az}\right)\right]
  \equiv\gN\n\left(\frac{\gN}{\az}\right).    \eeqno{nukmale}

\subsection{Form of the DML action  \label{dmlactionaq}}
In the DML, where $\f$ has SD $-\eta$, $
\hat\F$ has to have SD $-(\eta+3)$; so, it has to be a power in its argument
\beq   \hat\F\^D(u)=\a u^s,~~~~~~~s=\frac{\eta+3}{2(\eta+1)},  \eeqno{xv}
which gives the DML form of $\m\_G$:
 \beq   \m\_G\^D(x)=\a s x^{(1-\eta)/(1+\eta)},    \eeqno{xvi}
 which, specifically, leads to the asymptotic form of the potential and its gradient given by Eqs. (\ref{xiv}-\ref{xivas}), with $\gamma=(\a s)^{-(1+\eta)/2}$.

\par
The DML action is then
\beq  I\^D=-\frac{\a\az^{\frac{\eta-1}{\eta+1}}}{8\pi G}\int_{t_1}^{t_2} \frac{dt}{T}~\drt ~ |\gf|^{\frac{\eta+3}{\eta+1}}-\sum_i m_i\int_{t_1}^{t_2} \frac{dt}{T}~\f(\vvr_i(t))+\az^{-\eta}\sum_i m_i\U[\{\vvr_i(t)\}],  \eeqno{dmlaction}
and, in terms of the standardized quantities,
\beq  \tilde I\^D=-\frac{\a}{8\pi \azg} \int_{t_1}^{t_2} \frac{dt}{T}~\drt|\grad\tilde\f|^{\frac{\eta+3}{\eta+1}}-\sum_i m_i\int_{t_1}^{t_2} \frac{dt}{T}~\tilde\f(\vvr_i(t))+\sum_i m_i\U[\{\vvr_i(t)\}],   \eeqno{dmlactionstand}
where, as expected from the general argument, only $\azg$ appears.

\section{Summary and Discussion  \label{usessec}}
I have described families of NR theories that modify both the gravitational action and the inertial (kinetic) action of ND, without modifying the interaction between masses and the gravitational field. Excepting some examples, the modified actions were left in a rather general form, requiring only that the theory as a whole satisfies the basic tenets of MOND: Each of the two action terms, which depend on $\az$, is required to tend to its Newtonian form in the limit $\az\rar 0$. To boot, in the DML -- defined by $\az\rar\infty$, $G\rar 0$, with $\azg=\az G$ fixed -- the two actions are required to attain the same well-defined scaling behavior under spacetime scaling, in that they acquire the same scaling dimension, $-\eta$, under such scaling, which they have to share with the interaction term. At one end we have pure-MG theories, with $\eta=0$, and at the other, pure-MI theories with $\eta=1$. For each value of $\eta$, the scope of our treatment still covers a large class of possible actions of either MG or MI.
\par
The specific theories I discuss are not, in themselves, well motivated as candidate MOND theories, since the modifications of the two action terms do not appear to stem from some unified underlying principle beyond the fact that their DML expressions are required to have the same underlying symmetry.
My main purpose in propounding and discussing such theories is heuristic. Since we are still in the dark concerning the fundamentals of MOND, I wish to limelight the following: a. It may well be that a more fundamental understanding of MOND will point to modification of both gravity and inertia, unlike the existing MOND theories, which modify either of the two. b. In such theories neither of the two modified actions in itself leads to MOND phenomenology; it is only their combined effect that does. c. As expected, such theories do share the primary MOND predictions with other MOND theories that satisfy its basic axioms, but they can differ greatly in their predictions of some secondary phenomena, such as the workings of the MOND external-field effect, MOND effects in the solar system, etc.
\par
We can expect such differences based on the different ways interpolating functions emerge in such theories, compared with their roles in theories such as AQUAL and QUMOND, and based on the different ways the effects of the two modified actions have to combine to yield MOND phenomenology.

\appendix
\section{Theories anchored in QUMOND  \label{qumondapp}}
The actions of the family anchored in QUMOND (those for which the case $\eta=0$ gives QUMOND)
are
\beq  I=\frac{1}{8\pi G} \int_{t_1}^{t_2} \frac{dt}{T}~\drt \left[\az^2\hat\Q\left(\frac{|\grad\psi|^2}{\az^2}\right)-2\gf\cdot\grad\psi\right] -\sum_i m_i\int_{t_1}^{t_2} \frac{dt}{T}~\f(\vvr_i(t))+\sum_i m_i\IK[\{\vvr_i(t)\};\az].    \eeqno{qumondaction}
($\hat\Q$ does not, generally, have the same form as $\Q$ of QUMOND.)
\par
The particle equation of motion is the same as Eq. (\ref{muriso}).
Varying over $\psi$ shows it to be the Newtonian potential, $\psi=\fN$, as in standard QUMOND. Varying over $\f$ gives a QUMOND-like equation that is linear in $\f$
\beq \Delta\f =\div\left[\hat\n\left(\frac{|\gfN|}{\az}\right)\gfN\right], \eeqno{quqa}
with $\hat\n(y)=\hat\Q'(y^2)$.
\par
In the DML, all terms in the action must have the same well-defined SD, call it again $-\eta$. Then, as before, $\f$ has scaling dimension $-\eta$. Comparing the second and third terms in the action (\ref{qumondaction}), $\psi$ has to have SD equal -1, implying that in this limit $Q(z)\rar Q\^D(z)=\a z^s$, with $s=(\eta+3)/4$.
The standardized DML action $\tilde I\^D=\az^\eta I\^D$ is then given in terms of the standardized potentials $\tilde\f=\az^\eta\f$ and $\tilde\psi=\az\psi$ as
\beq  \tilde I\^D=\frac{1}{8\pi \azg} \int_{t_1}^{t_2} \frac{dt}{T}~\drt \left[\a|\grad\tilde\psi|^{(\eta+3)/2}-2\grad\tilde\f\cdot\grad\tilde\psi\right]-\sum_i m_i\int_{t_1}^{t_2} \frac{dt}{T}~\tilde\f(\vvr_i(t))+\sum_i m_i\U[\{\vvr_i(t)\}].    \eeqno{qumonddmlaction}

\end{document}